\documentclass[prl,aps,showpacs,preprintnumbers,twocolumn]{revtex4}
\usepackage{graphics,bm}
\begin{document}

\title{Many-body aspects of coherent atom-molecule oscillations}

\author{R.A. Duine}
\email{duine@phys.uu.nl}
\homepage{http://www.phys.uu.nl/~duine/bec}
\author{H.T.C. Stoof}
\email{stoof@phys.uu.nl}
\homepage{http://www.phys.uu.nl/~stoof}
\affiliation{Institute for Theoretical Physics,
         University of Utrecht, Leuvenlaan 4,
         3584 CE Utrecht, The Netherlands}
\date{February 18, 2003}

\begin{abstract}
We study the many-body effects on coherent atom-molecule
oscillations by means of an effective quantum field theory that
describes Feshbach-resonant interactions in Bose gases in terms
of an atom-molecule hamiltonian. We determine numerically the
many-body corrections to the oscillation frequency for various
densities of the atomic condensate. We also derive an analytic
expression that approximately describes both the
density and magnetic-field dependence of this frequency near the
resonance. We find excellent agreement with experiment.
\end{abstract}

\maketitle

{\it Introduction} --- One of the most remarkable applications of Feshbach
resonances in doubly spin-polarized alkali gases
\cite{stwalley,eite,inouye} is the observation of coherent atom-molecule
oscillations \cite{elisabeth2}. In this last experiment Donley {\it et
al.} used the Feshbach resonance at $B_0 \simeq 154.9$ Gauss in the $|
f=2;m_f=-2\rangle$ state of $^{85}$Rb to perform a Ramsey-type experiment,
consisting of two short pulses in the magnetic field towards resonance
separated by a longer evolution time. As a function of this evolution time
an oscillation in the number of condensate atoms was observed. Over the
investigated range of magnetic field during the evolution time, the
frequency of this oscillation agreed exactly with the molecular binding
energy found from a two-body coupled-channels calculation \cite{servaas},
indicating coherence between atoms and molecules.

Very recently, Claussen {\it et al.} have performed a similar
series of measurements over a larger range of magnetic fields
\cite{claussen2003}. It was found that close to resonance the
frequency of the oscillation deviates from the two-body molecular
binding energy, which indicates that many-body effects play an
important role in this regime. It is the main purpose of this
Letter to present the theory that explains this deviation
quantitatively.

The first mean-field theory for Feshbach-resonant interactions in
Bose-Einstein condensed gases is due to Drummond {\it et al.}
\cite{drummond1998} and Timmermans {\it et al.}
\cite{timmermans} and introduces the physical picture of an
interacting atomic condensate coupled to a noninteracting
molecular condensate. Although this theory contains the correct
resonant scattering amplitude for the atoms, it does not contain
the correct molecular binding energy. By studying the fluctuations
around this mean-field theory within the Hartree-Fock-Bogoliubov
approximation, it is possible to also incorporate the correct
binding energy \cite{servaas,mackie2002,kohler2002}. This comes
about because the latter approach explicitly contains a so-called
anomalous density, or pairing field, for the atoms which, after
elimination, leads to a shift in the coupling constants and the
molecular binding energy. Unfortunately, however, the elimination
does not lead to a proper renormalization of all the coupling
constants. In particular the interaction between condensate
atoms and non-condensate atoms, which is ultimately responsible
for the many-body corrections to the molecular binding energy
that are of interest to us here, is not correctly described
within the Hartree-Fock-Bogoliubov approximation.

Relying on the anomalous density for the description of the
molecular properties also makes the theory inapplicable above the
critical temperature for Bose-Einstein condensation, since the
anomalous density is proportional to the atomic condensate
density. To overcome all these problems, it is convenient to
formulate an effective quantum field theory that incorporates the
exact two-body physics not at the mean-field level but at the
quantum level and is hence applicable both above and below the
critical temperature. We have recently derived such an effective
quantum field theory by starting from the microscopic hamiltonian
for an atomic gas with a Feshbach resonance and explicitly summing all the ladder
diagrams \cite{rembert4}. Here, we apply the mean-field theory
for the Bose-Einstein condensed phase of the gas that results from this
effective quantum field theory, to the study of the recent
experiments by Claussen {\it et al.} \cite{claussen2003}.

{\it Atom-molecule coherence} --- The mean-field equations
consist of coupled equations for the macroscopic wave function
$\psi_{\rm a} ({\bf x},t)$, describing the atomic Bose-Einstein
condensate, and the macroscopic wave function $\psi_{\rm m} ({\bf
x}, t)$ that describes the condensate of bare molecules, i.e.,
with the appropriate bound-state wave function in the closed
channel potential of the Feshbach problem. The mean-field
equations for the coupled atom-molecule system are ultimately
given by
\begin{eqnarray}
\label{eq:mfe}
   i \hbar \frac{\partial \psi_{\rm a} ({\bf x},t)}{\partial t}
    &=& \left[-\frac{\hbar^2 {\bf \nabla}^2}{2m}
        +T^{\rm 2B}_{\rm bg} |\psi_{\rm a} ({\bf x},t)|^2
    \right]
      \psi_{\rm a} ({\bf x},t) \nonumber \\ &&+ 2 g  \psi_{\rm a}^* ({\bf x},t)
    \psi_{\rm m}({\bf x},t)~, \nonumber \\
   i \hbar \frac{\partial\psi_{\rm m} ({\bf x},t)}{\partial t}
    &=&\left[-\frac{\hbar^2 {\bf \nabla}^2}{4m}+ \delta (B(t))
  \right]  \psi_{\rm m} ({\bf x},t) +
  g \psi_{\rm a}^2 ({\bf x},t) \nonumber \\ && \hspace*{-0.7in}
  - g^2 \frac{m^{3/2}}{2 \pi \hbar^3} i \sqrt{i \hbar \frac{\partial}{\partial t}
    + \frac{\hbar^2 {\bf \nabla}^2}{4m} -2 \hbar \Sigma^{\rm HF}}
  \psi_{\rm m} ({\bf x},t)~,
\end{eqnarray}
where $T^{\rm 2B}_{\rm bg}=4 \pi a_{\rm bg} \hbar^2/m$ is the
off-resonant two-body T(ransition) matrix with $a_{\rm bg}$ the
off-resonant background scattering length and $m$ the mass of one
atom. The atom-molecule coupling $g$ is found from experiment by
adiabatically eliminating the molecular wave function and using
the fact that the resulting magnetic-field dependent scattering
length of the atoms must be equal to
\begin{equation}
\label{eq:ascatofb}
  a(B) = a_{\rm bg} \left( 1-\frac{\Delta B}{B-B_0} \right),
\end{equation}
with $\Delta B$ and $B_0$ the experimental width and position of
the resonance, respectively. This procedure results in $g=\hbar \sqrt{2 \pi
a_{\rm bg} \Delta B  \Delta \mu /m}$, where we have made use of
the fact that the detuning of the bare molecular state is given
by $\delta(B) = \Delta \mu (B-B_0)$ with $\Delta \mu \simeq -2.2 \mu_{\rm
B}$ for $^{85}$Rb \cite{servaas} and $\mu_{\rm B}$ the Bohr
magneton.

The mean-field equation for the molecular condensate contains a
fractional derivative, corresponding to the retarded self energy
of the molecules. In momentum and frequency space, this
self energy reads $\hbar \Sigma^{(+)}_{\rm m} ({\bf k},\omega)=-
(g^2 m^{3/2}/2 \pi \hbar^3) i
  \sqrt{\hbar \omega - \hbar^2{\bf k}^2/4m -2 \hbar \Sigma^{\rm HF}}$
\footnote{The full expression for the self energy of the
molecules is given by $\hbar\Sigma^{(+)}_{\rm m} ({\bf k},\omega)= - (g^2
m^{3/2}/2 \pi \hbar^3) i
  \sqrt{z} (1+i a_{\rm bg}\sqrt{m z}/\hbar)^{-1}$,
with $z = \hbar \omega - \hbar^2{\bf k}^2/4m -2 \hbar \Sigma^{\rm
HF}$. Close to resonance only the square-root in the numerator is
important. We will use this approximation throughout the paper.}.
The square-root behavior is a result of the Wigner threshold law
for the decay of a molecule with total energy $\hbar \omega$ and
center-of-mass momentum $\hbar {\bf k}$ into
the two-atom continuum. Due to the mean-field interaction of the
noncondensed atoms with the condensate, the decaying molecule has
to overcome a mean-field barrier given by $2 \hbar \Sigma^{\rm
HF}$. Here, $\hbar \Sigma^{\rm HF}$ denotes the Hartree-Fock
self energy for the noncondensed atoms. Neglecting the momentum
dependence of this self energy, it is given by
\begin{eqnarray} \label{eq:sigmahf}
   &&\hbar \Sigma^{\rm HF} =
 2 n_{\rm a} \Bigl( T^{\rm 2B}_{bg} \nonumber \\
 &&+\frac{2 g^2}{\hbar \Sigma^{\rm HF}+\mu
 -\delta (B) - g^2 \frac{m^{3/2}}{2 \pi \hbar^3} \sqrt{\hbar
 \Sigma^{\rm HF}-\mu}} \Bigr)~,
\end{eqnarray}
where $n_{\rm a}=|\psi_{\rm a}|^2$ is the density of the atomic
condensate, $\mu$ its chemical potential, and we have used that a
collision between a condensed atom and a noncondensed atom has a
mean-field shift of $\mu+\hbar \Sigma^{\rm HF}$. Far from resonance the
energy-dependence of the interactions can be safely ignored and the
Hartree-Fock self energy becomes equal to $8 \pi a(B) \hbar^2  n_{\rm
a}/m$, as expected.

The molecular binding energy is given by the pole of the
molecular propagator at zero momentum, which from
Eq.~(\ref{eq:mfe}) is seen to be given by
\begin{equation}
\label{eq:gmkw}
  G_{\rm m}^{(+)} ({\bf 0},\omega)=
    \frac{\hbar}{\hbar \omega+i0   -\delta (B)
     + \frac{g^2 m^{3/2}}{2 \pi \hbar^3} i
     \sqrt{\hbar \omega - 2\hbar \Sigma^{\rm HF}}}~.
\end{equation}
In the limit of vanishing condensate density $n_{\rm a}$ and for
negative detuning, it has a pole at
\begin{equation}
\label{eq:bse}
  \epsilon_{\rm m} (B)= \delta (B) + \frac{g^4 m^3}{8 \pi^2 \hbar^6}
   \left[\sqrt{1-\frac{16 \pi^2 \hbar^6}{g^4 m^3} \delta (B)} -1\right].
\end{equation}
Close to the resonance it thus follows that \mbox{$\epsilon_{\rm m} (B) =
-\hbar^2/[m a(B)^2]$}, which is the correct molecular binding energy in
vacuum \cite{servaas}. Note that in the absence of the molecular self
energy the binding energy would be equal to the detuning, which is
incorrect close to resonance. Moreover, the residue of the pole is given
by
\begin{eqnarray}
\label{eq:factorz}
  Z(B)&=&\left[1-\frac{\partial \Sigma^{(+)}_{\rm m}({\bf 0},\omega)}
                      {\partial \omega}\right]^{-1}
  \nonumber \\ &=& \left[1 + \frac{g^2 m^{3/2}}{4 \pi \hbar^3
\sqrt{|\epsilon_{\rm m}(B)|}}\right]^{-1}
\end{eqnarray}
and always smaller than one. Physically, the latter can be understood from
the fact that the dressed
molecular bound state near the Feshbach resonance is given by
\begin{eqnarray}
\label{eq:wavefctmol}
  | \chi_{\rm m}; {\rm dressed} \rangle&=&
                 \sqrt{Z(B)}|  \chi_{\rm m} ; {\rm bare} \rangle
                 \nonumber \\
                &&+ \int \frac{d {\bf k}}{(2 \pi)^3} C({\bf k})
                           | {\bf k}, -{\bf k};{\rm open} \rangle ~,
\end{eqnarray}
where the coefficient $C({\bf k})$ denotes the amplitude of the dressed
molecular state to be in the open channel of the Feshbach problem and the
two atoms having momenta ${\bf k}$ and $-{\bf k}$, respectively. They are
normalized as $\int d {\bf k} |C({\bf k})|^2/(2 \pi)^3  = 1-Z(B)$. The
dressed molecular state therefore only contains with an amplitude $\sqrt{Z(B)}$ the
bare molecular state $|\chi_{\rm m};{\rm bare} \rangle$. Close to
resonance we have that $Z(B) \ll 1$, whereas it approaches one far off
resonance. With respect to this remark it is important to note that the
result of the Hartree-Fock-Bogoliubov theory for
the density of the molecular condensate should be multiplied by a factor
$1/Z(B) \gg 1$ to obtain the density of real dressed molecules, since in
this theory always the density of bare molecules is calculated
\cite{rembert5,braaten2003}.

\begin{figure}
\includegraphics{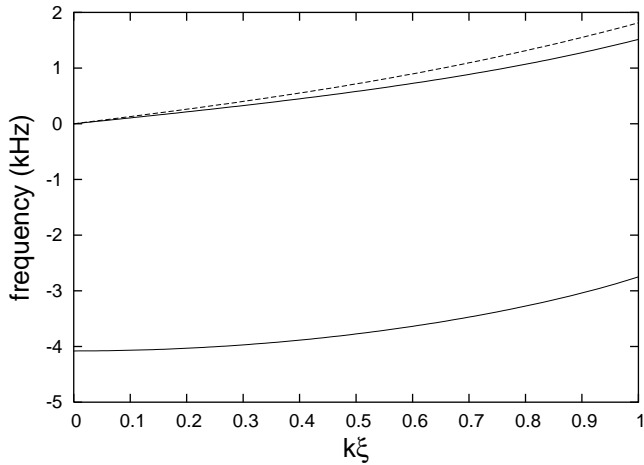}
\caption{\label{fig:fig1}
   Dispersion relation for the collective modes of the atom-molecule system
   for an atomic condensate density of $n_{\rm a}=2 \times 10^{12}$
   cm$^{-3}$, at a magnetic field of $B=156$ G. The momentum is indicated
   in units of $1/\xi$, where $\xi = 1/\sqrt{16 \pi a n_{\rm a}}$ is the 
   coherence length. The upper solid line shows
   the gapless dispersion relation for phonon-like excitations. The lower
   solid line indicates the dispersion for atom-molecule oscillations. The
   dashed line shows the effective Bogoliubov dispersion.
   }
\end{figure}

{\it Collective modes} --- To study the many-body effects on the frequency
of the coherent atom-molecule oscillations it is important to realize that
these oscillations are in fact a collective mode where the atomic
condensate density oscillates out-of-phase with the molecular condensate
density. It is thus worthwhile to study the collective modes of the
mean-field equations in Eq.~(\ref{eq:mfe}) and look for solutions of the
form
\begin{eqnarray}
\label{eq:linearization}
  \psi_{\rm m} ({\bf x},t)\!&=&\!\left[ \psi_{\rm m}
    +u'_{\bf k} e^{-i \omega t + i {\bf k} \cdot {\bf x}}
    +{v'_{\bf k}}^* e^{+i \omega t -i {\bf k} \cdot {\bf x}}\right]e^{-i 2 \mu
 t/\hbar}~,
 \nonumber \\
\psi_{\rm a} ({\bf x},t)\!&=&\!\left[ \psi_{\rm a} +u_{\bf k}
e^{-i \omega t + i {\bf k} \cdot {\bf x}}
 + v_{\bf k}^* e^{+i \omega t -i {\bf k} \cdot {\bf x}}\right]e^{-i \mu
 t/\hbar}~.
\end{eqnarray}
After substitution into the mean-field equations in
Eq.~(\ref{eq:mfe}) the eigenmodes are found by diagonalizing the
resulting $4 \times 4$ matrix. This yields a dispersion relation
with two branches, one corresponding to the gapless Bogoliubov
modes and one that corresponds to the atom-molecule oscillations.
The zero-momentum part of the latter corresponds to the
experimentally observed frequency of the coherent atom-molecule
oscillations. Note that for this calculation the evaluation of
the molecular self energy can be performed exactly because every
term in the right-hand side of Eq.~(\ref{eq:linearization}) is an
eigenfunction of the operator under the square root.

In Fig.~\ref{fig:fig1} we show the dispersion relations found by means of
the above procedure, for a fixed atomic condensate density of $n_{\rm a}=2
\times 10^{12}$ cm$^{-3}$ at a fixed magnetic field of $B=156$ G.
Physically, the upper branch corresponds to the phonon-like excitations in
the atomic condensate and the lower branch to coherent atom-molecule
oscillations. The dashed line denotes the Bogoliubov dispersion for the
scattering length $a(B)$. At low momenta the phonon branch corresponds
with the Bogoliubov dispersion, as expected. At higher momenta the
dispersion starts to deviate from the Bogoliubov result, due to the energy
and momentum dependence of the resonant interactions that reduce the
scattering amplitude. The dispersion of the lower branch obeys
$\hbar\omega_{\bf k} \simeq -\hbar\omega_{\rm J} + \hbar^2{\bf k}^2/4m$,
where $\hbar\omega_{\rm J}$ is the  Josephson frequency that is observed
by Claussen {\it et al}. in their Ramsey experiment. This dispersion is
negative due to the fact that we are dealing with a metastable situation.
For negative detuning the true ground state contains almost
all atoms in the form of molecules. 
\begin{figure}
\includegraphics{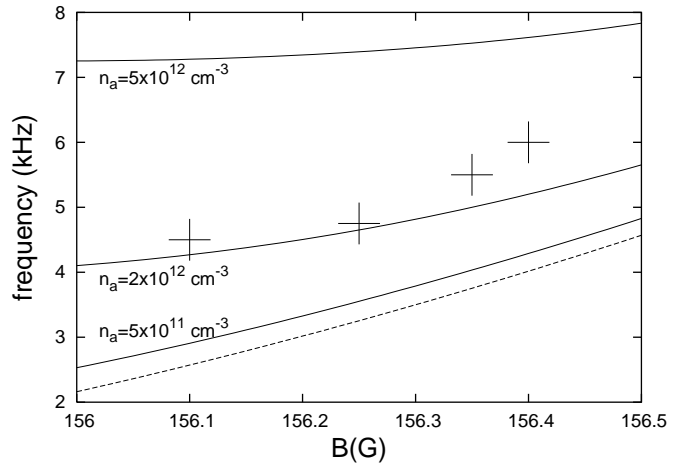}
\caption{\label{fig:fig2}
   Frequency of coherent atom-molecule oscillations. The solid lines show
   the frequency as a function of the magnetic field for three different
   densities of the atomic condensate. The dashed line shows the two-body
   result for the molecular binding energy. The experimental points are
   taken from Ref.~\cite{claussen2003}.
   }
\end{figure}

\begin{figure}
\includegraphics{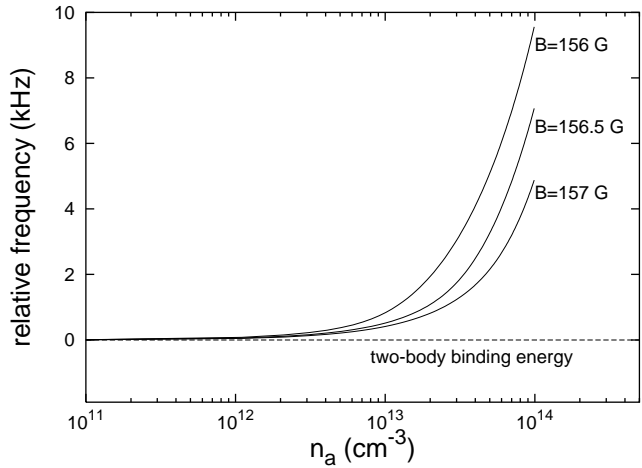}
\caption{\label{fig:fig3}
   Frequency of coherent atom-molecule oscillations as a function of the
   atomic condensate density, relative to the two-body binding energy. The
   solid lines show the result for different magnetic fields.
   }
\end{figure}

In Fig.~\ref{fig:fig2} the results are shown for the 
frequency of the atom-molecule oscillations as a function of the
magnetic field, for three different densities. Clearly, for
increasing density the frequency starts to deviate from the
two-body result. In addition, we show in Fig.~\ref{fig:fig3} the
frequency of the atom-molecule oscillations relative to the
two-body binding energy, as a function of the atomic condensate
density. The calculation is performed in this case for several
values of the magnetic field. Both the magnetic field and the
atomic density dependence of the frequency can be understood
as follows. We first observe that for the range of magnetic
fields and atomic densities that are explored experimentally, the
difference between the energy of the dressed molecular state and
the threshold of the two-particle continuum is almost independent
of density and equal to $\epsilon_{\rm m}(B)$. It is interesting
to note that only this difference is independent of density,
whereas both quantities individually show a substantial
mean-field shift of about $2 \hbar \Sigma^{\rm HF}$ 
\cite{rembert4}. As a result of the cancellation,
however, also the wave-function renormalization factor $Z$ is
almost density independent and equal to $Z(B)$. Expressing our
mean-field equations in Eq.~(\ref{eq:mfe}) in terms of the
condensate wave function for the dressed molecules by replacing
$\psi_{\rm m}({\bf x},t)$ by $\sqrt{Z(B)} \psi_{\rm m}({\bf
x},t)$, we see that the coupling between the atomic condensate
and the dressed molecular condensate is reduced by a factor of
$\sqrt{Z(B)} \ll 1$ \cite{rembert5}. We thus expect the frequency to obey
\begin{equation}
\label{eq:josephson}
  \hbar\omega_{\rm J} \simeq
     \sqrt{16 Z(B) g^2 n_{\rm a} + (\epsilon_{\rm m} (B))^2}~.
\end{equation}
This analytic expression indeed turns out to give a first approximation to
the deviation of the frequency from the two-body result.

In order to confront our results with the experimental data we have to
realize that the experiments are performed in a magnetic trap. Taking only
the ground-states into account for both the atomic and the molecular
condensates, this implies effectively that the atom-molecule coupling $g$
is reduced by an overlap integral. Hence we define the effective
homogeneous  condensate density by means of $n_{\rm a} = N_{\rm a} \left[
\int d {\bf x} \phi_{\rm a}^2 ({\bf x} ) \phi_{\rm m} (\bf x) \right]^2$,
where $N_{\rm a}$ denotes the number of condensed atoms and $\phi_{\rm a}
({\bf x} )$ and $\phi_m (\bf x)$ denote the atomic and molecular ground
state wave function, respectively. For the experiments of Claussen {\it et
al.} this results in an effective density of $n_{\rm a}\simeq 2 \times
10^{12}$ cm$^{-3}$ \cite{claussen2003}. Fig.~\ref{fig:fig2} clearly shows
an excellent agreement with the experimentally observed frequency for this
density.

It is important to note that there are two hidden assumptions in
the above comparison. First, we have used that the dressed
molecules are trapped in the same external potential as the
atoms. This is not obvious because the bare molecular state
involved in the Feshbach resonance is high-field seeking and
therefore not trapped. However, Eq.~(\ref{eq:factorz}) shows that
near resonance almost all the amplitude of the dressed molecule
is in the low-field seeking open channel and its magnetic moment
is therefore almost equal to twice the atomic magnetic moment.
Second, we have determined the frequency of the coherent
atom-molecule oscillations in equilibrium. In contrast, the
observed oscillations in the number of condensate atoms is clearly
a nonequilibrium phenomenon. This is, however, expected not to
play an important role because the Ramsey-pulse sequence is
performed on such a fast time scale that the response of the
condensate wave function can be neglected. 

{\it Conclusions} --- With the linear-response calculation presented in
this Letter we have obtained excellent agreement with the experimental
results on the frequency of coherent atom-molecule oscillations. The next
step is a more detailed understanding of other quantities that are of
interest for the two-pulse experiments \cite{elisabeth2,claussen2003}.
This includes a quantitative study of the number of condensed atoms as a
function of time that goes beyond the linear approximation discussed here.
Experimentally an overall decay of the number of condensed atoms is
observed and also that the oscillations have a finite damping rate. Both
effects increase as one approaches the resonance. We expect that an
important contribution to  these effects is due to the so-called
rogue-dissociation process \cite{mackie2002}, which has its physical origin
in the decay of a molecule into two noncondensed atoms that in the
center-of-mass system have opposite momenta ${\bf k}$ and $-{\bf k}$,
respectively. In equilibrium this process is always forbidden due to
energy conservation, since the molecular state lies below the two-atom
continuum threshold. However, in the nonequilibrium setting of the
experiments of interest, it may occur due to the fact that the detuning is
strongly time-dependent. In the mean-field equations in Eq.~(\ref{eq:mfe})
it is the fractional derivative term corresponding to the self energy of
the molecules that automatically incorporates the rogue-dissociation
effect. A full numerical simulation of these equation is challenging due
to the fact that we are essentially dealing with  a term that is nonlocal
in time. Nevertheless, work in this direction is in progress and will be
reported in a future publication.

\end{document}